\documentclass[12pt]{iopartmod1}
\usepackage{color, graphicx}
\usepackage{bm}
\usepackage{iopams}
\usepackage{accents}

\font\smalltesti=cmmi8 at 8.0pt
\font\smallseveni=cmmi7 at 5.83pt
 1  

\begin{document}

\title[Interpretation of twisting type N vacuum solutions with cosmological constant]
{Interpretation of twisting type N vacuum solutions with cosmological constant}

\author{Xuefeng Zhang and Daniel Finley}

\address{Department of Physics and Astronomy, University of New Mexico, Albuquerque, NM 87131 USA}
\eads{\mailto{zxf@unm.edu} and \mailto{finley@phys.unm.edu}}

\begin{abstract}
We investigate a new class of twisting type N vacuum solutions with
nonzero (positive) cosmological constant $\Lambda$ by studying the
equations of geodesic deviations along the privileged radial
timelike geodesics, generalizing J. Bi\v{c}\'{a}k and J.
Podolsk\'{y}'s results on non-twisting type N solutions. It is shown
that these twisting radiative spacetimes can be interpreted as exact
transverse gravitational waves propagating in the de-Sitter
universe, with a distinctive feature that all the wave amplitudes
are proportional to $\Lambda$ (dark energy/matter source coupling).
Moreover, we demonstrate the cosmic no-hair conjecture in these
spacetimes and discuss their Killing horizons.
\end{abstract}


\submitto{\CQG}

\maketitle


\section{Introduction}
\label{introduction}

In a recent paper \cite{Zhang12a}, we presented a new class of
twisting type N vacuum solutions of the Einstein equations with
nonzero cosmological constant $\Lambda$. These type N solutions
admit a twisting congruence of shearfree and null geodesics aligned
with the unique quadruple principal null direction. They were shown
subject to a rather simple-looking second-order nonlinear ODE as
imposed by the field equations for type N. Various special and
series solutions were found or constructed from this ODE. In this
paper, we move on to discuss their physical meanings and show that
these new exact solutions can serve as models for the behavior of
gravitational waves in cosmology.

Certain aspects of these solutions, such as the conformal factor,
conformal infinities, periodicity, etc., can be found in the general
discussion on algebraically special twisting spacetimes by Hill and
Nurowski \cite{Hill08}. Here we focus on the local interpretation
associated with the equation of geodesic deviation. Similar analysis
has been done for non-twisting type N solutions with $\Lambda$,
i.e., Kundt class and Robinson-Trautman class,  by Bi\v{c}\'{a}k and
Podolsk\'{y} \cite{Bicak99} (see also \cite{Podolsky98,
Podolsky04}). However, a complication for our solutions is that the
nonzero twist is associated with certain non-integrability, and
therefore causes a lack of 2-dimensional wave surfaces in the
spacetimes that always exist in non-twisting solutions. Hence due to
extra cross terms in the metric, the coordinate basis adopted in
\cite{Bicak99} cannot be directly used in our more complicated
twisting solutions, and we decide that it would be much more
convenient to use non-holonomic bases instead, and therefore we
generalize all derivations to such bases. In fact, we find that all
major results involving the equations of geodesic deviation remain
the same for this generalization.

In the next section, we review the solutions discovered in
\cite{Zhang12a} and present their Levi-Civita connection 1-forms and
the only non-vanishing Weyl scalar, $\Psi_4$. Then in Section 3, we
write down the general geodesic equations in the non-holonomic basis
and use Killing symmetries to simplify them. In Sections 4 and 5, a
frame for a physical observer and an associated null tetrad basis
are constructed along arbitrary timelike geodesics, with respect to
which the relative motion of test particles is studied. Conditions
are given to determine those geometrically privileged geodesics
along which the observer's frame can be parallel-transported. Hence
in Section 6, these privileged geodesics are identified as the
radial geodesics and determined explicitly. Their properties are
discussed in Section 7, which are all pointing to a Killing horizon.
In Section 8, we calculate wave amplitudes along radial timelike
geodesics and demonstrate that the observer sees gravitational waves
decaying exponentially fast, which agrees with the cosmic no-hair
conjecture. At the end, we make comments on the appearance of
$\Lambda$ as a proportionality factor in the waves amplitudes, which
we find quite unusual.

\section{The twisting type N vacuum spacetimes with nonzero $\Lambda$}

With the real coordinate system $x^\alpha=(x, J, u, r)$, the metric
found in \cite{Zhang12a} can be written as
\begin{equation} \label{metric}
 \mathbf{g} = 2 \left( \boldsymbol{\omega}^1 \boldsymbol{\omega}^2 + \boldsymbol{\omega}^3 \boldsymbol{\omega}^4 \right),
\end{equation}
where the null tetrad is given by
\begin{equation} \label{tetrad}
 \eqalign{
 \boldsymbol{\omega}^1 = {R}\,\rmd\zeta, \qquad \boldsymbol{\omega}^2 = {R}\,{\rmd\bar\zeta},  \cr
 \boldsymbol{\omega}^3 = {R}\,\lambda, \qquad \ \boldsymbol{\omega}^4 = {R}
 \left(\rmd r + W\rmd\zeta + \bar W \rmd\bar\zeta + H\lambda \right), }
\end{equation}
with real-valued $P=P(J)>0$ and $P' \equiv \rmd P/\rmd J$ such that
\begin{equation} \label{components}
 \eqalign{
 \rmd\zeta = \rmd x + \frac{\rmi}{P} \rmd J, \qquad
 \lambda = \frac{ \rmd u + 2 L \rmd x}{ -P \partial_J L}, \\
 L = -\rme^{-C_1 x} \int \! \frac{1}{P} \exp\! \left( \int \! \frac{P'-2\Lambda J}{2P} \rmd J \right) \rmd J, \\
 {R} = \frac{\sqrt{P}}{2 \cos\case{r}{2}}, \qquad -\pi<r<\pi, \\
 W = \frac{1}{2} \left( \frac{1}{2} P' + \Lambda J + \rmi C_1 \right) (\mathrm{e}^{-\rmi
 r}+1), \qquad
 H = -\frac{1}{6} \Lambda P \cos r. }
\end{equation}
Here $C_1$ is an arbitrary real parameter and the function $P(J)$
must satisfy
\begin{equation} \label{2ODE}
 P'' = - \frac{(P' + 2 \Lambda J)^2}{2P} - \frac{2 C_1^2}{P} - \frac{10}{3} \Lambda,
\end{equation}
as required by the vacuum Einstein field equations with real
arbitrary constant $\Lambda$. By the transformation $J=w/\Lambda$,
$P=g(w)/\Lambda>0$, with $\Lambda\neq 0$, the above equation can be
put into a $\Lambda$-independent form
\begin{equation} \label{2ODE1}
 g'' = - \frac{(g' + 2w)^2}{2g} - \frac{2 C_1^2}{g} - \frac{10}{3}.
\end{equation}
These ODEs allow solutions for either $\Lambda>0$ or $\Lambda<0$,
though the construction of the explicit general solution, if
possible, still remains an open problem. (See \cite{Zhang12a} for
examples of special solutions and series solutions.)

Now we introduce the non-holonomic null basis
$\{\mathbf{e}_\alpha\} = \{\mathbf{e}_1, \mathbf{e}_2, \mathbf{e}_3, \mathbf{e}_4 \}$
dual to the tetrad \eref{tetrad}:
\begin{equation} \label{basis}
 \eqalign{
 \mathbf{e}_1 = \frac{1}{{R}}\,(\partial - W \partial_r), \qquad
 \mathbf{e}_2 = \frac{1}{{R}}\,(\bar\partial - \bar W \partial_r), \cr
 \mathbf{e}_3 = \frac{1}{{R}}\,(\partial_0 - H \partial_r), \qquad
 \mathbf{e}_4 = \frac{1}{{R}}\,\partial_r, }
\end{equation}
where we define
\begin{equation} \label{partial}
 \eqalign{ \partial = \partial_\zeta - L \partial_u, \qquad
 \partial_\zeta = \case{1}{2}\,(\partial_x - \rmi P \partial_J), \cr
 \partial_0 = \rmi (\bar\partial L - \partial \bar L) \partial_u = -P(\partial_J L) \partial_u. }
\end{equation}
In particular, the vector field $\mathbf{e}_4$ is tangent to a
twisting congruence of shearfree null geodesics, and is also aligned
with the quadruple principal null direction of the metric. To ensure
a nonzero twist along this congruence (also for the basis
$\{\mathbf{e}_\alpha\}$ to be valid), it is required that
\begin{equation}
 \rmi (\bar\partial L - \partial \bar L) = -P(\partial_J L) \neq 0.
\end{equation}
With all said, we present components of the Levi-Civita connection
1-forms $\boldsymbol{\Gamma}^\lambda_{\phantom{\lambda}\mu} =
\Gamma^\lambda_{\phantom{\lambda}\mu\nu} \boldsymbol{\omega}^\nu$
calculated from Cartan's structure equations
$\rmd\boldsymbol{\omega}^\lambda +
\boldsymbol{\Gamma}^\lambda_{\phantom{\lambda}\mu} \wedge
\boldsymbol{\omega}^\mu = 0$ for the null tetrad \eref{tetrad}:
\begin{equation} \label{Gamma}
 \eqalign{
 \Gamma_{121} = -\overline{\Gamma}_{122} = \frac{\rmi}{8 {R}} \left[(\rme^{-\rmi r}+1) P'
 + 2 (\rme^{-\rmi r}-1) (\Lambda J + \rmi C_1) \right], \cr
 \Gamma_{123} = -\frac{\rmi}{12 {R}} \Lambda P (2 + \cos r) , \qquad \Gamma_{124} = -\frac{\rmi}{2{R}}, \cr
 \Gamma_{231} = \frac{1}{12 {R}} \Lambda P \left(\rmi\rme^{-\rmi r} - \tan\frac{r}{2} + 2\rmi \right),  \cr
 \Gamma_{233} = \frac{\rmi}{24 {R}} \Lambda P (\rme^{-\rmi r}+1) (P' + 2\Lambda J - 2\rmi C_1), \cr
 \Gamma_{241} = \frac{\rmi\, \rme^{-\rmi r/2}}{2 {R} \cos \case{r}{2}} \cr
 \Gamma_{341} = \overline{\Gamma}_{342} = -\frac{\rmi}{8 {R}} \left[(\rme^{-\rmi r}-1) P'
 + 2 (\rme^{-\rmi r}+1) (\Lambda J + \rmi C_1) \right], \cr
 \Gamma_{343} = \frac{1}{12 {R}} \Lambda P (2+\cos r) \tan\frac{r}{2}, \qquad
 \Gamma_{344} = \frac{1}{2 {R}} \tan\frac{r}{2}, \cr
 \Gamma_{232} = \Gamma_{234} = \Gamma_{242} = \Gamma_{243} = \Gamma_{244} = 0. }
\end{equation}
Those unlisted components can be obtained by either
$\boldsymbol{\Gamma}_{\mu\nu} = -\boldsymbol{\Gamma}_{\nu\mu}$ or
complex conjugation on \eref{Gamma} which interchanges the indices
$1\leftrightarrow2$ and leaves 3 and 4 unchanged, e.g.,
$\Gamma_{214}=\overline{\Gamma}_{124}$,
$\Gamma_{211}=\overline{\Gamma}_{122}$. The only non-vanishing Weyl
scalar is
\begin{equation} \label{Psi4}
 \fl \eqalign{
 \Psi_4 = C_{3232} = -\frac{\Lambda}{3} \left[ \Lambda J P' -\case{2}{3} \Lambda P + 2 \Lambda^2 J^2 - 4 C_1^2
 -2 \rmi C_1 \left(P' + 3\Lambda J \right)\right] \rme^{-\rmi r/2} \cos^3\frac{r}{2} \cr
 \qquad \qquad \ \ = -\frac{\Lambda}{3} \left[ w g' -\case{2}{3} g + 2 w^2 - 4 C_1^2
 -2 \rmi C_1 \left(g' + 3w \right)\right] \rme^{-\rmi r/2} \cos^3\frac{r}{2}, }
\end{equation}
which, being proportional to $\Lambda$, requires $\Lambda\neq 0$ for
type N solutions. In addition, the metric has the following two
Killing vectors \cite{Zhang12b}
\begin{equation} \label{KV}
 \partial_u, \qquad \partial_x - C_1 u\, \partial_u,
\end{equation}
which we will use to simplify the geodesic equations.

\section{Geodesic equations}

Using the non-holonomic basis \eref{basis}, we consider a freely
falling test particle (observer) with the four-velocity
\begin{equation*}
 u^\alpha = (u^1, u^2, u^3, u^4), \qquad u^1=\overline{u^2},
\end{equation*}
along an arbitrary timelike geodesic such that
\begin{equation} \label{length}
 \mathbf{u}\cdot\mathbf{u}=2(u^1 u^2+u^3 u^4)=\epsilon
\end{equation}
with $\epsilon=-1$ (also, $\epsilon=0$ if one considers null
geodesics). From the fact that $\mathbf{u}=u^\alpha
\mathbf{e}_\alpha=\dot{x}^\mu
\partial_{x^\mu}$ as expressed in non-holonomic and coordinate
bases, we obtain
\begin{equation} \label{xjdot}
 \eqalign{
 \dot{x} = \frac{1}{2{R}}\,(u^1+u^2), \qquad \dot{J} = -\frac{\rmi P}{2{R}}\,(u^1-u^2), \cr
 \dot{u} = -\frac{1}{{R}}\left[ L(u^1 + u^2) + P(\partial_J L) u^3 \right], \cr
 \dot{r} = -\frac{1}{{R}}\,(W u^1 + \bar{W} u^2 + H u^3 - u^4). }
\end{equation}
with $\cdot\equiv \rmd/\rmd\tau$ and $\tau$ the proper time. The
geodesic equations read
\begin{equation} \label{GE}
 \eqalign{
 0 = \frac{\rmd u^1}{\rmd\tau} + \Gamma_{2\mu\nu} u^\mu u^\nu, \qquad
 0 = \frac{\rmd u^2}{\rmd\tau} + \Gamma_{1\mu\nu} u^\mu u^\nu, \cr
 0 = \frac{\rmd u^3}{\rmd\tau} + \Gamma_{4\mu\nu} u^\mu u^\nu, \qquad
 0 = \frac{\rmd u^4}{\rmd\tau} + \Gamma_{3\mu\nu} u^\mu u^\nu, }
\end{equation}
from which one can verify, using \eref{Gamma}, that
$\frac{\rmd}{\rmd\tau}(u^1u^2+u^3u^4)=0$ (hence, the length
\eref{length} is constant along geodesics). For a Killing vector
$\xi^\mu$, the product $u_\mu \xi^\mu$ is a conserved quantity along
geodesics with the four-velocity $u^\mu$. Hence from \eref{KV}, we
find
\begin{equation} \label{CQ}
 \eqalign{
 C_2 = \frac{{R}}{-P\partial_J L} \left( H u^3 + u^4 \right),  \\
 C_3 = {R} \left[ u^1 + u^2 + (W+\bar{W}) u^3 \right] + 2 C_2 L - C_2 C_1 u(\tau), }
\end{equation}
where $C_2$ and $C_3$ are real constants. One can show that the
above two simpler equations are indeed first integrals of the system
\eref{GE}. However, with an implicit $P(J)$ generally satisfying
\eref{2ODE}, these geodesic equations are still very complicated to
solve directly for $x^\mu(\tau)$ without additional assumptions.

\section{Geodesic deviation}

The equation of geodesic deviation reads
\begin{equation} \label{deviation}
 \frac{D^2 Z^\mu}{\rmd \tau^2} = -R^{\mu}_{\phantom{\mu}\alpha\beta\gamma} u^\alpha Z^\beta u^\gamma,
\end{equation}
where $\mathbf{u}=\rmd\mathbf{x}/\rmd\tau=u^\alpha
\mathbf{e}_\alpha$, $\mathbf{u}\cdot\mathbf{u}=-1$ as introduced
before, and $\mathbf{Z}(\tau)$ is the displacement vector. Note that
\eref{deviation} is still valid despite the use of non-holonomic
bases \cite{Misner73}. Following the construction\footnote{We will
use the indices $\{1, 2, 3, 4\}$ for basis elements instead of $\{0,
1, 2, 3\}$ adopted in \cite{Bicak99}, i.e., changing $0$ to $4$, and
changing the ordering.} in \cite{Bicak99}, we set up the observer's
frame $\{\mathbf{e}_{(\alpha)}\}$ along the geodesic with
$\mathbf{e}_{(4)}=\mathbf{u}$ and spacelike orthonormal vectors
$\{\mathbf{e}_{(1)}, \mathbf{e}_{(2)}, \mathbf{e}_{(3)}\}$ in the
local hypersurface orthogonal to $\mathbf{u}$, i.e.,
$\mathbf{e}_{(\alpha)} \cdot \mathbf{e}_{(\beta)} = g_{\mu\nu}
e^\mu_{(\alpha)} e^\nu_{(\beta)} = \eta_{(\alpha)(\beta)} =
\mathrm{diag}(1,1,1,-1)$. The dual basis is
$\mathbf{e}^{(4)}=-\mathbf{u}$ and
$\mathbf{e}^{(i)}=\mathbf{e}_{(i)}, i=1,2,3$. Then we can project
\eref{deviation} onto the observer's frame:
\begin{equation} \label{deviation1}
 \ddot{Z}^{(\alpha)} \equiv \mathbf{e}^{(\alpha)} \cdot \frac{D^2\mathbf{Z}}{\rmd\tau^2}
 = e^{(\alpha)}_\mu \frac{D^2 Z^\mu}{\rmd\tau^2}
 = -R^{(\alpha)}_{\phantom{(\alpha)}(4)(\beta)(4)} Z^{(\beta)}
\end{equation}
with $Z^{(\beta)} = \mathbf{e}^{(\beta)}\cdot \mathbf{Z} =
e^{(\beta)}_\mu Z^{\mu}$ and $R_{(\alpha)(4)(\beta)(4)} =
e^\mu_{(\alpha)} u^\nu e^\gamma_{(\beta)} u^\delta
R_{\mu\nu\gamma\delta}$. Now we introduce the second null basis
$\{\mathbf{e}_{\hat{\alpha}}\} = \{\mathbf{e}_{\hat{1}},
\mathbf{e}_{\hat{2}}, \mathbf{e}_{\hat{3}}, \mathbf{e}_{\hat{4}}\} =
\{\mathbf{m}, \bar{\mathbf{m}}, \mathbf{l}, \mathbf{k}\}$ associated
with the observer's frame:
\begin{equation} \label{frame}
 \eqalign{
 \mathbf{m} = \frac{1}{\sqrt{2}} \left(\mathbf{e}_{(1)} + \rmi \mathbf{e}_{(2)}\right), \qquad
 \bar{\mathbf{m}} = \frac{1}{\sqrt{2}} \left(\mathbf{e}_{(1)} - \rmi \mathbf{e}_{(2)}\right), \cr
 \mathbf{l} = \frac{1}{\sqrt{2}} \left(\mathbf{u} - \mathbf{e}_{(3)}\right), \qquad
 \mathbf{k} = \frac{1}{\sqrt{2}} \left(\mathbf{u} + \mathbf{e}_{(3)}\right). }
\end{equation}
One can check that all derivations in Section II of \cite{Bicak99}
also hold for our non-holonomic basis $\{\mathbf{e}_\alpha\}$. Here
we only quote those results relevant to our purpose. To begin with,
all test particles should be synchronized so that $Z^{(4)} = 0$
(they always stay in the same local hypersurface). For type N
spacetimes, the rest of \eref{deviation1} reads
\begin{equation} \label{GD}
 \eqalign{
 \ddot{Z}^{(1)} = \left(\frac{\Lambda}{3} - \mathcal{A}_+ \right) Z^{(1)} + \mathcal{A}_\times Z^{(2)}, \cr
 \ddot{Z}^{(2)} = \left(\frac{\Lambda}{3} + \mathcal{A}_+ \right) Z^{(2)} + \mathcal{A}_\times Z^{(1)}, \cr
 \ddot{Z}^{(3)} = \frac{\Lambda}{3} Z^{(3)}
 }
\end{equation}
with wave amplitudes of the two polarization modes given by
\begin{equation} \label{amp}
 \eqalign{
 \mathcal{A}_+ = \frac{1}{2} \mathcal{R}e \hat\Psi_4, \qquad
 \mathcal{A}_\times = \frac{1}{2} \mathcal{I}m \hat\Psi_4, \qquad
 \hat\Psi_4 = C_{\alpha\beta\gamma\delta} l^\alpha \bar{m}^\beta l^\gamma \bar{m}^\delta. }
\end{equation}
Here $\hat\Psi_4$ is calculated in the second null basis
\eref{frame}, hence different from, but related to $\Psi_4$ in
\eref{Psi4}. Assuming that the observer's frame
$\{\mathbf{e}_{(\alpha)}\}$ is parallel-transported along
$\mathbf{u}$, i.e., $D\mathbf{e}^{(\alpha)}/\rmd\tau=0$, then we
have
$\ddot{Z}^{(\alpha)}=D^2(\mathbf{e}^{(\alpha)}\cdot\mathbf{Z})/\rmd\tau^2=\rmd^2
Z^{(\alpha)}/\rmd\tau^2$, which makes \eref{GD} easier to solve.

\section{Parallel-transported frames}

Given a radiative spacetime with a principal null direction
$\mathbf{k}$ and an observer's four-velocity $\mathbf{u}$, we can
construct the observer's frame $\{\mathbf{e}_{(\alpha)}\}$ according
to \eref{frame} together with the following proposition .

\textbf{Proposition 1.} \cite{Bicak99} Let $\mathbf{u}$ be the
observer's four-velocity and $\mathbf{k}$ be the null vector
(principal null directions) that satisfy
$\mathbf{k}\cdot\mathbf{u}=-\frac{1}{\sqrt{2}}$ . Then there is a
unique spacelike vector $\mathbf{e}_{(3)} =
\sqrt{2}\mathbf{k}-\mathbf{u}$. Another null vector $\mathbf{l}$ is
given by $\mathbf{l}=\sqrt{2}\mathbf{u} - \mathbf{k}$ such that
$\mathbf{l}\cdot\mathbf{k}=-1$. The only remaining freedoms are
rotations in the transverse plane $(\mathbf{e}_{(1)},
\mathbf{e}_{(2)})$ perpendicular to $\mathbf{e}_{(3)}$.

With the null basis $\{\mathbf{e}_1, \mathbf{e}_2, \mathbf{e}_3, \mathbf{e}_4\}$,
the metric \eref{metric} takes the very simple form
\begin{equation} \label{matrix}
 g_{\mu\nu} = \left( \begin{array}{cccc}
 0 & 1 & 0 & 0 \\
 1 & 0 & 0 & 0 \\
 0 & 0 & 0 & 1 \\
 0 & 0 & 1 & 0
 \end{array}\right).
\end{equation}
We take the vector field $\mathbf{k}$ to be aligned with the
quadruple principal null direction $\mathbf{e}_4$, i.e.,
\begin{equation*}
 k^\mu = (0, 0, 0, k^4),
\end{equation*}
and recall that $u^\mu = (u^1, u^2, u^3, u^4)$, $u^1=\overline{u^2}$
in the basis $\{\mathbf{e}_\alpha\}$. The interpretation null basis
described in \eref{frame} and Proposition 1 has the form
\begin{equation} \label{ibasis}
 \eqalign{
 m^\mu = \left(0, -1, 0, \frac{u^1}{u^3} \right), \qquad
 \bar{m}^\mu = \left(-1, 0, 0, \frac{u^2}{u^3} \right), \cr
 l^\mu = \left(\sqrt{2}u^1, \sqrt{2}u^2, \sqrt{2}u^3, \sqrt{2}u^4 + \frac{1}{\sqrt{2}u^3} \right), \cr
 k^\mu = \left(0, 0, 0, -\frac{1}{\sqrt{2}u^3} \right), }
\end{equation}
which is unique up to rotations $\mathbf{m} \rightarrow
\rme^{\rmi\theta}\mathbf{m}$ and trivial reflections. The
corresponding orthonormal frame in the basis $\{\mathbf{e}_\alpha\}$
is given by
\begin{equation} \label{orthnframe}
 \eqalign{
 e^\mu_{(1)} = \frac{1}{\sqrt{2}}\left(-1, -1, 0, \frac{u^1+u^2}{u^3} \right), \cr
 e^\mu_{(2)} = \frac{1}{\sqrt{2}}\left(-\rmi, \rmi, 0, \frac{u^1-u^2}{\rmi u^3} \right), \cr
 e^\mu_{(3)} = -\left(u^1, u^2, u^3, u^4+\frac{1}{u^3} \right), \cr
 e^\mu_{(4)} = u^\mu = (u^1, u^2, u^3, u^4). }
\end{equation}
One can check that these expressions are consistent with Eqn. (18)
and (19) of \cite{Bicak99} when applied to \eref{matrix} with matrix
elements rearranged accordingly.

In general, the frames $\{\mathbf{e}_{(\alpha)}\}$ and
$\{\mathbf{e}_{\hat\alpha}\}$ cannot be parallel-transported along
the geodesic with $\mathbf{u}=\mathbf{e}_{(4)}$. We can use the
following proposition to single out those geometrically privileged
geodesics along which the interpretation frames are indeed
parallel-transported. The detailed proof can be found in
\cite{Podolsky93} with no difficulty to be generalized for the
non-holonomic basis \eref{basis} with the metric \eref{matrix}.

\textbf{Proposition 2.} \cite{Bicak99,Podolsky93} Given a geodesic
with the tangent vector $u^\mu=(u^1, u^2, u^3, u^4)$ in the
spacetime \eref{matrix}, the interpretation null basis \eref{ibasis}
and the orthonormal frame \eref{orthnframe} are parallel-transported
along the geodesic if
\begin{equation} \label{PT1}
 0 = \Gamma^1_{\phantom{1}4\mu} u^\mu
\end{equation}
and
\begin{equation} \label{PT2}
 \dot\vartheta_{\|}(\tau) = \rmi\, \Gamma^2_{\phantom{2}2\mu} u^\mu
\end{equation}
where $\vartheta_{\|}$ is the rotation angle for $\mathbf{m}
\rightarrow \mathbf{m}_{\|} = \rme^{\rmi\vartheta_{\|}} \mathbf{m}$
in order that $D\mathbf{m}_{\|}/\rmd\tau=0$.

\section{Privileged geodesics}

Now we apply the above results to our twisting type N spacetimes
with $\Lambda$. First recall that
$\Gamma_{242}=\Gamma_{243}=\Gamma_{244}=0$. Then the condition
\eref{PT1} and its complex conjugate are tantamount to
\begin{equation}
 u^1 = u^2 = 0,
\end{equation}
which, by \eref{xjdot}, leads to
\begin{equation}
 \dot{x} = \dot{J} = 0.
\end{equation}
Along such \emph{radial} geodesics with fixed $x$ and $J$,
the geodesic equations are quite simplified with the first two of \eref{GE} given by
\begin{equation} \label{GE12}
 0 = \frac{\rmi \Lambda {R}}{24 P (\partial_J L)^2}\, (\rme^{-\rmi r}+1) (P' + 2\Lambda J - 2\rmi C_1)\, \dot{u}^2
\end{equation}
and its complex conjugate.
The rest of \eref{GE} are given by
\begin{eqnarray}
 0 = \ddot{u} + \left( \frac{\Lambda}{6 \partial_J L} \tan\frac{r}{2} \right)\! \dot{u}^2, \label{GE3} \\
 0 = \ddot{r} + \tan\frac{r}{2}\, \left(\dot{r}^2
 - \frac{\Lambda}{3\,\partial_J L}\, \dot{r}\dot{u}
 - \frac{\Lambda^2 \cos r}{18(\partial_J L)^2}\, \dot{u}^2 \right). \label{GE4}
\end{eqnarray}
The conserved quantities \eref{CQ}, i.e., first integrals of (\ref{GE3}, \ref{GE4}), read
\begin{eqnarray}
 0 = \dot{r} + \frac{\Lambda \cos r}{3\partial_J L}\, \dot{u} + 4 C_2 (\partial_J L) \cos^2\frac{r}{2}, \label{CQ1} \\
 0 = \frac{(W+\bar{W})}{\partial_J L}\, \dot{u} + 4(C_1 C_2 u - 2C_2 L + C_3) \cos^2\frac{r}{2} \label{CQ2},
\end{eqnarray}
in addition to the invariant length \eref{length}
\begin{equation} \label{length1}
 0 = \frac{\epsilon}{2} - \left( \frac{\Lambda \cos r}{24 (\partial_J L)^2 \cos^2\frac{r}{2}}\, \dot{u}
 + C_2 \right)\! \dot{u}.
\end{equation}
Note that $\partial_J L$ and $L$ above are \emph{constant} for fixed $x$ and $J$
and that $W$ is given by \eref{components}.

Now we proceed to solve the system (\ref{GE12}--\ref{GE4}).
First note that the equation \eref{GE12} gives rise to the following two possibilities.

\emph{Case 1.} We have $\dot{u}=0$. Then \eref{length1} requires
$\epsilon=0$. This is the case corresponding to null geodesics along
principle null directions. The geodesic equation \eref{GE4}
immediately gives us
\begin{equation*}
 0 = \ddot{r} + \dot{r}^2 \tan\frac{r}{2},
\end{equation*}
which has the general solution
\begin{equation*}
 r = 2 \arctan(A\tau+B)
\end{equation*}
with $A$, $B$ arbitrary real constants of integration.

\emph{Case 2.} Assuming $\dot{u}\neq 0$ and a given solution
$P_0(J)$ to \eref{2ODE}, we have, for \eref{GE12} to hold,
\begin{equation} \label{PG}
 0 = \frac{\rmd P_0(J)}{\rmd J} + 2\Lambda J, \qquad C_1=0.
\end{equation}
The first equation above, with its right-hand side being a function
of $J$, shall fix the value of $J$ which we call $J_0$. In fact,
combining \eref{PG} with \eref{2ODE}, we know that $J_0$ is also the
point at which the second derivative $P_0''(J)$ reaches its maximum
value $-10\Lambda/3$. Hence for a given $P_0(J)$, we only have
limited choices of $J$ for the privileged geodesics described in
Proposition 2. Nonetheless, the coordinate $x$ can take on any
arbitrary constant value which we denote as $x=x_0$. In what
follows, such $P=P_0(J_0)$ and $x=x_0$ will always be assumed.

To solve the system (\ref{GE3}, \ref{GE4}), one can first solve
\eref{GE3} for $r$ and then combine the result with the first
integral \eref{CQ1} so as to obtain a third-order ODE for $u(\tau)$
alone
\begin{equation}
 0 = \dot{u}\, \dddot{u} - \ddot{u}^2 - \frac{C_2 \Lambda}{3}\, \dot{u}^3 - \frac{\Lambda^2}{36C_L^2}\, \dot{u}^4,
 \qquad \Lambda\neq 0.
\end{equation}
Thus for this equation, we obtain in the \emph{real} domain the
following two different ways\footnote{In the complex domain, the
expressions \eref{usol1} and \eref{usol2} are in fact equivalent. To
see this, one can separate a constant from $E_1$ and then use the
addition theorem of $\tanh$ on the left-hand side of \eref{usol1}
and compare it with \eref{usol2}. However in the real domain, this
transformation from \eref{usol1} to \eref{usol2}, for their full
ranges of real integration constants, cannot be achieved without
generally going into the complex domain (e.g., note that
$\mathrm{arctanh}(x)$ is complex for $x>1$). For certain limited
ranges of real integration constants, the solutions \eref{usol1} and
\eref{usol2} may represent the same solutions. } to represent its
general solution both of which we find suitable for timelike
geodesics with $\Lambda>0$ (cf. \eref{epsilon}):
\begin{eqnarray}
 \tanh\! \left(\frac{\Lambda}{12 C_L}\, (u-E_1) \right)
 = \pm( \rme^{A_1\tau+B_1} + D_1), \label{usol1} \\
 \tanh\! \left(\frac{\Lambda}{12 C_L}\, (u-E_2) \right)
 = \sqrt{1+D_2^2}\, \tanh\! \left(\frac{A_2\tau+B_2}{2} \right) + D_2, \label{usol2}
\end{eqnarray}
with $C_L \equiv \partial_J L(J_0)\neq 0$ and real integration
constants $A_1, B_1, D_1, E_1$ ($C_2=A_1 D_1/2C_L$) and $A_2, B_2,
D_2, E_2$ ($C_2=-A_2/2C_L\!\sqrt{1+D_2^2}\,$). By the equation
\eref{GE3}, these $u(\tau)$'s yield respectively
\begin{eqnarray}
 r(\tau) = \pm 2 \arctan\! \left(
 - \frac{1}{2}\, \rme^{-(A_1\tau+B_1)} (1-D_1^2)
 - \frac{1}{2}\, \rme^{A_1\tau+B_1} \right), \label{rsol1} \\
 r(\tau) = -2 \arctan\! \left(
 \frac{D_2^2}{\sqrt{1+D_2^2}} \sinh(A_2\tau+B_2) + D_2 \cosh(A_2\tau+B_2) \right).
 \label{rsol2}
\end{eqnarray}
One can check that the expressions (\ref{usol1}, \ref{rsol1}) and
(\ref{usol2}, \ref{rsol2}) all constitute general solutions to
(\ref{GE3}, \ref{GE4}) in the real domain. In particular when
$D_2=0$ in (\ref{usol2}, \ref{rsol2}), one has a special solution
\begin{equation} \label{spesol}
 u(\tau) = \frac{6C_L}{\Lambda}\,(A_2\tau+B_2)+E_2, \qquad r(\tau)=0,
\end{equation}
which is not included in the solution (\ref{usol1}, \ref{rsol1}).
From \eref{length1}, we obtain for both sets of solutions
\begin{equation} \label{epsilon}
 \epsilon = -\frac{3A_{1,2}^2}{\Lambda},
\end{equation}
with $\epsilon=-1$ for timelike geodesics. In addition, the equation
\eref{CQ2} yields trivially $C_3=2C_2 L(J_0)$.

\section{Radial timelike geodesics and Killing horizon}

We continue to study timelike geodesics given by (\ref{usol1},
\ref{rsol1}) and (\ref{usol2}, \ref{rsol2}). For simplicity, we only
consider the more physically relevant situation with $\Lambda>0$.
Hence from \eref{epsilon}, one has
\begin{equation} \label{A12}
 A_{1,2} = \pm \sqrt{\frac{\Lambda}{3}}, \qquad \Lambda>0
\end{equation}
for timelike geodesics. Note that the range of the hyperbolic
function tanh on the left-hand sides of \eref{usol1} and
\eref{usol2} is limited to the interval $[-1,1]$, while the
right-hand sides are not. Therefore the solution $u(\tau)$ with its
$\tau$ restricted by the reality condition may generally reach
infinity at some \emph{finite} proper time. Nonetheless, whenever
this happens at the critical value $\tau=\tau_c$ such that
\begin{equation*}
 \rme^{A_1\tau_c + B_1} + D_1 = \pm 1, \qquad D_1^2 \neq 1,
\end{equation*}
or
\begin{equation*}
 \sqrt{1+D_2^2}\, \tanh\! \left(\frac{A_2\tau_c+B_2}{2} \right) + D_2 = \pm 1, \qquad D_2\neq 0,
\end{equation*}
it always corresponds to
\begin{equation}
 r(\tau_c) = 2\arctan(\mp 1)= \mp \frac{\pi}{2},
\end{equation}
which are in fact very special hypersurfaces in the spacetime. We
can see their significance from the metric \eref{metric} with
constant $x$ and $J$
\begin{equation} \label{metricur}
 \widetilde{\mathbf{g}} = -\frac{1}{2 C_L \cos^2\frac{r}{2}}\, \rmd u \left( \rmd r +
 \frac{\Lambda}{6C_L} \cos r\, \rmd u \right), \qquad -\pi<r<\pi,
\end{equation}
which indicates that the Killing vector $\partial_u$ is timelike for
$-\pi/2<r<\pi/2$ but spacelike for $-\pi<r<-\pi/2$ and
$\pi/2<r<\pi$, and \emph{null} at $r=\pm\pi/2$. Thus that $u(\tau)$
diverges at $r=\pm\pi/2$ under the metric \eref{metricur} is very
similar to that of the Schwarzschild solution in the
Eddington-Finkelstein coordinates near $r=2m$, the fact of which
suggests a \emph{Killing horizon}\footnote{Due to cross terms like
$\rmd\zeta \rmd u$ in the metric with the current coordinate system,
it is not at all clear how to show $r=\pm\pi/2$ an actual horizon
for non-radial causal geodesics. Hence we will not dwell on this
issue.} at $r=\pm\pi/2$. Inside the region $r=\pm\pi/2$, the
spacetime is \emph{stationary} with $u$ being a time coordinate
(e.g., \eref{spesol}). Particularly in the weak field limit
$\Psi_4\rightarrow 0$ with finite $\Lambda>0$ \cite{Zhang12a}, one
can expect that $r=\pm\pi/2$ approaches the cosmological horizon of
the de Sitter universe.

\section{Wave amplitudes}

Now we calculate the wave amplitudes $\mathcal{A}_+$ and
$\mathcal{A}_\times$ in \eref{GD}. The null bases
$\{\mathbf{e}_\alpha\}$ and $\{\mathbf{e}_{\hat{\alpha}}\}$ (cf.
\eref{basis} and \eref{frame}) are related by the Lorentz
transformation
\begin{equation}
 \eqalign{
 \mathbf{e}_4 = A\, \mathbf{e}_{\hat{4}}, \cr
 \mathbf{e}_1 = \rme^{\rmi\theta} \mathbf{e}_{\hat{2}} + \bar{B} \mathbf{e}_{\hat{4}}, \cr
 -\mathbf{e}_3 = A^{-1}( \mathbf{e}_{\hat{3}} + B\rme^{\rmi\theta} \mathbf{e}_{\hat{2}}
 + \bar{B} \rme^{-\rmi\theta} \mathbf{e}_{\hat{1}} + B\bar{B} \mathbf{e}_{\hat{4}} ), }
\end{equation}
with $A=-\sqrt{2}u^3$, $B=-\sqrt{2}u^1$ and $\theta=\pi$. The Weyl
scalar $\Psi_4$ transforms as \cite{Stephani03}
\begin{equation}
 \hat{\Psi}_4 = A^2 \bar\Psi_4
 = \frac{4 \Lambda^2 P_0(J_0)}{9}\, (u^3)^2\, \rme^{\rmi r/2} \cos^3\frac{r}{2} \propto \Lambda,
\end{equation}
where $u^3 = -{R} (\dot{u} + 2 L \dot{x}) /(P\partial_J L)$ from
\eref{xjdot} (also, use the re-scaled function $g(w)$ from
\eref{2ODE1} to see the proportionality to $\Lambda$). Hence we know
\begin{equation}
 \eqalign{
 \mathcal{A}_+ = \frac{1}{2} \mathcal{R}e \hat\Psi_4
 = \frac{\Lambda^2}{18 C_L^2}\, \dot{u}^2 \cos^2\frac{r}{2}, \cr
 \mathcal{A}_\times = \frac{1}{2} \mathcal{I}m \hat\Psi_4
 = \frac{\Lambda^2}{18 C_L^2}\, \dot{u}^2 \sin\frac{r}{2}\, \cos\frac{r}{2}.
 }
\end{equation}
Substituting either (\ref{usol1}, \ref{rsol1}) or (\ref{usol2},
\ref{rsol2}) for $u(\tau)$ and $r(\tau)$ above with \eref{A12}, we
find that as long as the limit $u(+\infty)$ exists (e.g., when
$A_1<0$ in \eref{usol1}) and $r(+\infty)=\pm\pi$, the amplitudes
behave like
\begin{equation}
 \mathcal{A}_+ \sim \Lambda \exp \left(-4 \sqrt{\frac{\Lambda}{3}}\ \tau \right), \qquad
 \mathcal{A}_\times \sim \Lambda \exp \left(-3 \sqrt{\frac{\Lambda}{3}}\ \tau \right).
\end{equation}
as the proper time $\tau \rightarrow +\infty$. This means that the
gravitational waves are decaying exponentially fast with the
spacetime locally approaching the de-Sitter universe. Hence the
cosmic no-hair conjecture \cite{Maeda89} is demonstrated with these
radial geodesics. Furthermore, due to the proportionality to
$\Lambda$, the polarization modes $\mathcal{A}_+$ and
$\mathcal{A}_\times$ cannot be generally separated from the
isotropic background represented by $\Lambda/3$ in \eref{GD}, when
it comes to consider their local effects.

\section{Concluding remarks}

Like their non-twisting counterparts \cite{Bicak99}, the twisting
spacetimes described by (\ref{metric}-\ref{2ODE}) bear a similar
local interpretation as exact transverse gravitational waves in the
de-Sitter universe. However, the \emph{essential}\footnote{Note that
the appearance of $\Lambda$ in $\Psi_4$ (cf. \eref{Psi4}), hence
also in the wave amplitudes, is a direct consequence of solving the
vacuum Einstein equations $R_{ij}=\Lambda g_{ij}$ for type N rather
than being artificially introduced.} nonzero requirement of the
cosmological constant $\Lambda$ in our radiative solutions (wave
amplitudes proportional to $\Lambda$; for $\Lambda=0$, spacetimes
becoming flat), to our best knowledge, has not been seen in any
other radiative exact solution (e.g., Kundt class and
Robinson-Trautman class). This distinctive feature of these
spacetimes suggests that the gravitational waves they represent may
not be considered as being generated by usual astronomical sources
which always radiate regardless of the \emph{background} $\Lambda$
being zero or not. Moreover, from the linear theory of gravitational
waves with cosmological constant, one can indeed identify extra
freedom in the higher-order terms of approximate solutions caused by
the $\Lambda$ parameter \cite{Bernabeu11}, which is consistent with
our observation from the exact theory. In fact, according to
\cite{Bernabeu11}, this extra freedom in the wave solutions has been
described as being ``coupling to matter sources with a strength
proportional to the cosmological constant itself'', though the
actual physics behind this \emph{ad hoc} coupling is quite unknown.
Altogether, both theories suggest a quite different role that the
cosmological constant (or perhaps, dark energy) may play in the
process of gravitational radiations, other than simply the
``inactive'' de-Sitter background as it is generally thought of.
Since in the inflationary epoch, the effect of $\Lambda$ on the
cosmology was much greater than at present, we expect that these
radiative solutions might be of more relevance to primordial
gravitational waves from the Big Bang.






\end{document}